# Minimizing the Continuous Diameter when Augmenting Paths and Cycles with Shortcuts[*]

Jean-Lou De Carufel[1], Carsten Grimm[2,3], Anil Maheshwari[2], and Michiel Smid[2]

1  School of Electrical Engineering and Computer Science, University of Ottawa
   800 King Edward Avenue, Ottawa, Ontario, K1N 6N5, Canada
2  School of Computer Science, Carleton University
   1125 Colonel By Drive, Ottawa, Ontario, K1S 5B6, Canada
3  Institut für Simulation und Graphik, Otto-von-Guericke-Universität Magdeburg
   Universitätsplatz 2, D-39106 Magdeburg, Germany
   carsten.grimm@ovgu.de

## Abstract

We seek to augment a geometric network in the Euclidean plane with shortcuts to minimize its continuous diameter, i.e., the largest network distance between any two points on the augmented network. Unlike in the discrete setting where a shortcut connects two vertices and the diameter is measured between vertices, we take all points along the edges of the network into account when placing a shortcut and when measuring distances in the augmented network.

We study this network augmentation problem for paths and cycles. For paths, we determine an optimal shortcut in linear time. For cycles, we show that a single shortcut never decreases the continuous diameter and that two shortcuts always suffice to reduce the continuous diameter. Furthermore, we characterize optimal pairs of shortcuts for convex and non-convex cycles. Finally, we develop a linear time algorithm that produces an optimal pair of shortcuts for convex cycles. Apart from the algorithms, our results extend to rectifiable curves.

Our work reveals some of the underlying challenges that must be overcome when addressing the discrete version of this network augmentation problem, where we minimize the discrete diameter of a network with shortcuts that connect only vertices.

## 1 Introduction

The minimum-diameter network augmentation problem is concerned with minimizing the largest distance between two vertices of an edge-weighted graph by introducing new edges as shortcuts. We study the this problem in a continuous and geometric setting where the network is a geometric graph embedded into the Euclidean plane, the weight of a shortcut is the Euclidean distance of its endpoints, and shortcuts can be introduced between any two points along the network that may be vertices or points along edges.

As a sample application, consider a network of highways where we measure the distance between two locations in terms of the travel time. An urban engineer might want to improve the worst-case travel time along a highway or along a ring road by introducing shortcuts. Our work advises where these shortcuts should be build. For example, we show where to find the best shortcut for a highway and we show that a single shortcut never improves the worst-case travel time of a ring road. Our continuous perspective on network augmentation reflects that a shortcut road could connect any two locations along a highway.

---

[*] This work was partially supported by NSERC.

## 1.1 Preliminaries

A *network* is an undirected graph that is embedded into the Euclidean plane and whose edges are weighted with their Euclidean length. We say a point $p$ lies on a network $G$ and write $p \in G$ when there is an edge $e$ of $G$ such that $p$ is a point along the embedding of $e$. A point $p$ on an edge $e$ of length $l$ subdivides $e$ into two sub-edges lengths $(1 - \lambda) \cdot l$ and $\lambda \cdot l$ for some value $\lambda \in [0, 1]$. By specifying the points on $G$ in terms of their relative position (expressed by $\lambda$) along their containing edge, we avoid any ambiguity in case of crossings.

The *network distance* between two points $p$ and $q$ on a network $G$ is the length of a weighted shortest path from $p$ to $q$ in $G$. We denote the network distance between $p$ and $q$ by $d_G(p, q)$ and we omit the subscript when the network is understood. The largest network distance between any two points on $G$ is the *continuous diameter* of $G$, denoted by $\mathrm{diam}(G)$, i.e., $\mathrm{diam}(G) = \max_{p,q \in G} d_G(p, q)$. The term *continuous* distinguishes this notion from the *discrete diameter* that measures the largest network distance between any two vertices.

We denote the Euclidean distance between $p$ and $q$ by $|pq|$. A line segment $pq$, with $p, q \in G$ is a *shortcut* for $G$ when $|pq| < d_G(p, q)$. We augment a network $G$ with a shortcut $pq$ as follows. We introduce new vertices at $p$ and at $q$ in $G$, subdividing their containing edges, and we add a edge from $p$ to $q$ of length $|pq|$. The resulting network is denoted by $G + pq$. We seek to minimize the continuous diameter of a network by introducing shortcuts.

## 1.2 Related Work

In the discrete abstract setting, we consider an abstract graph $G$ with unit weights and ask whether we can decrease the discrete diameter of $G$ to at most $D$ by adding at most $k$ edges. For any fixed $D \geq 2$, this problem is NP-hard [1, 6, 8], has parametric complexity W[2]-hard [3, 4], and remains NP-hard even if $G$ is a tree [1]. For an overview of the approximation algorithms in terms of both $D$ and $k$ refer, for instance, to Frati *et al.* [3].

In the discrete geometric setting, we consider a geometric graph, where a shortcut connects two vertices. Große *et al.* [5] are the first to consider diameter minimization in this setting. They determine a shortcut that minimizes the discrete diameter of a path with $n$ vertices in $O(n \log^3 n)$ time. The spanning ratio of a geometric network, i.e., the largest ratio between the network distance and the Euclidean distance of any two points, has been considered as target function for edge augmentation, as well. For instance, Farshi *et al.* [2] compute a shortcut that minimizes the spanning ratio in $O(n^4)$ time while Luo and Wulff-Nilsen [7] compute a shortcut that maximizes the spanning ratio in $O(n^3)$ time.

## 1.3 Structure and Results

Our results concern networks that are paths, cycles, and convex cycles. Figures 1 and 2 illustrate examples of optimal shortcuts for paths and cycles. In Section 2, we develop an algorithm that produces an optimal shortcut for a path with $n$ vertices in $O(n)$ time. In Section 3, we show that for cycles a single shortcut never suffices to reduce the diameter and that two shortcuts always suffice. We characterize pairs of optimal shortcuts for convex and non-convex cycles. Based on this characterization, we develop an algorithm in Section 4 that determines an optimal pair of shortcuts for a convex cycle with $n$ vertices in $O(n)$ time.

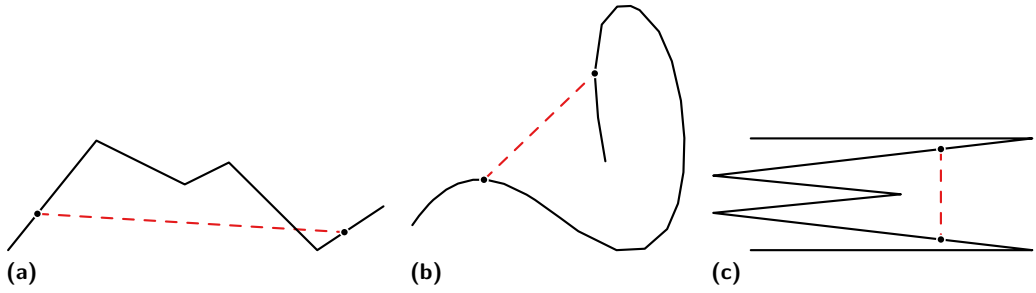

**Figure 1** Examples for paths with an optimal shortcut.

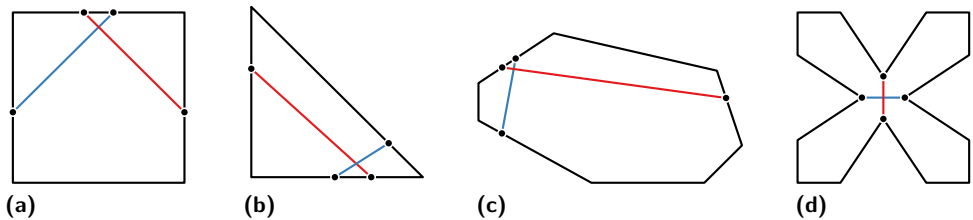

**Figure 2** Examples for cycles with optimal pairs of shortcuts.

## 2 Shortcuts for Paths

Consider a polygonal path $P$ in the plane. We seek a shortcut $pq$ for a path $P$ that minimizes the continuous diameter of the augmented path $P + pq$, i.e.,

$$\mathrm{diam}(P + pq) = \min_{a,b \in P} \mathrm{diam}(P + ab) = \min_{a,b \in P} \max_{u,v \in P+ab} d_{P+ab}(u,v) \ .$$

The following notation is illustrated in Figure 3. Let $s$ and $e$ be the endpoints of $P$ and let $p$ be closer to $s$ than $q$ along $P$, i.e., $d(s,p) < d(s,q)$. For $a, b \in P$, let $P[a,b]$ denote the sub-path from $a$ to $b$ along $P$, and let $C(p,q)$ be the simple cycle in $P + pq$.

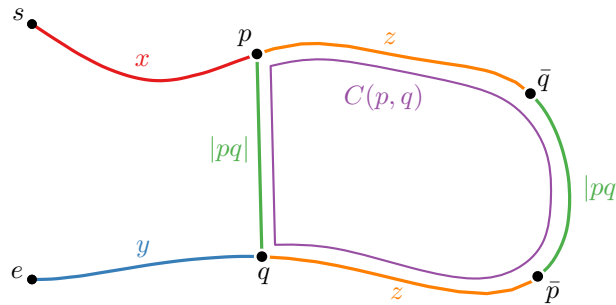

**Figure 3** Augmenting a path $P$ with a shortcut $pq$. The shortcut creates a cycle $C(p,q)$ with the sub-path from $p$ to $q$ along $P$. The farthest point from $p$ on this cycle is $\bar{p}$ and $\bar{q}$ is farthest from $q$ on $C(p,q)$. The distance $d(\bar{q}, \bar{p})$ between $\bar{q}$ and $\bar{p}$ along $P$ matches the Euclidean distance between $p$ and $q$, because of the following. When we move a point $g$ from $p$ to $q$ along the shortcut $pq$, then the farthest point $\bar{g}$ form $g$ along $C(p,q)$ moves from $\bar{p}$ to $\bar{q}$ traveling the same distance as $g$, i.e., $|pq|$.

▶ **Lemma 2.1.** *Let $pq$ be a shortcut for $P$. Every continuous diametral path in $P + pq$ contains an endpoint of $P$, except when the shortcut connects the endpoints of $P$.*

**Proof.** Let $P$ be a polygonal path with endpoints $s$ and $e$, and let $pq$ be a shortcut for $P$.

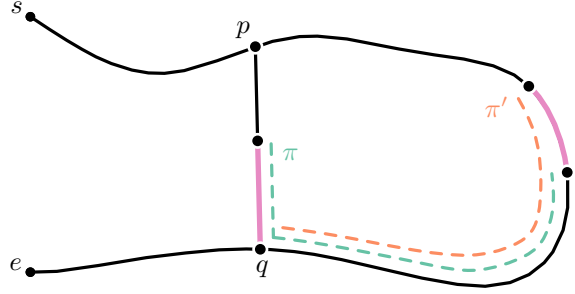

**Figure 4** We can shift any path $\pi$ with endpoints on $C(p, q)$ until one of them coincides with $q$. The shifted path $\pi'$ can be extended by the sub-path $P[e, q]$, i.e., $\pi$ is not continuous diametral.

Let $\pi$ be a path along $P + pq$. When an endpoint of $\pi$ lies on $P[s, p]$ or on $P[e, q]$, we can create a longer path by extending $\pi$ to $s$ or to $e$. Suppose both endpoints of $\pi$ lie on the simple cycle $C(p, q)$. As illustrated in Figure 4, we can move the endpoints of $\pi$ at the same speed counter-clockwise along $C(p, q)$ until one them coincides with $p$ or with $q$. The resulting path $\pi'$ has the same length as $\pi$ and can be extended by $P[s, p]$ or $P[e, q]$ creating a longer path than $\pi$, unless $s = p$ and $e = q$. Therefore, for every shortcut $pq$ other than $se$, every continuous diametral path in $P + pq$ must contain and endpoint of $P$. ◀

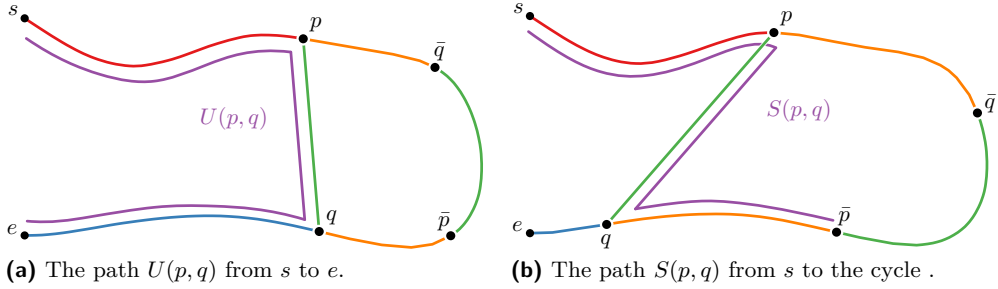

(a) The path $U(p, q)$ from $s$ to $e$.  (b) The path $S(p, q)$ from $s$ to the cycle .

**Figure 5** Two candidate diametral paths in $P + pq$, namely the shortest path connecting $s$ and $e$ in (a) and a path from $s$ via $p$ to the farthest point from $p$ on the cycle $C(p, q)$ in (b). For the latter, there is a second path $S'(p, q)$ of the same length traversing $C(p, q)$ in the other direction.

According to Lemma 2.1, we have the following three candidates for continuous diametral paths in the augmented network $P + pq$, two of which are illustrated in Figure 5.
1. The path $U(p, q)$ from $s$ to $e$ via the shortcut $pq$,
2. the path $S(p, q)$ from $s$ to the farthest point from $s$ on $C(p, q)$, and
3. the path $E(p, q)$ from $e$ to the farthest point from $e$ on $C(p, q)$.

Let $\bar{p}$ be the farthest point from $p$ on $C(p, q)$, and let $\bar{q}$ be the farthest point from $q$ on $C(p, q)$. Furthermore, let $\delta(p, q) := \frac{d(p,q) - |pq|}{2}$ denote the slack between $p$ and $\bar{q}$ (and symmetrically between $\bar{p}$ and $q$) along $C(p, q)$. With this notation, we have

$$d(p, \bar{p}) = d(q, \bar{q}) = \frac{|C(p,q)|}{2} = \frac{d(p,q) + |pq|}{2} = \frac{d(p,q) - |pq|}{2} + |pq| = \delta(p, q) + |pq| \ ,$$

and we can express the lengths of $U(p, q)$, $S(p, q)$, and $E(p, q)$ as follows.

$$|U(p, q)| = d(s, p) + |pq| + d(q, e)$$
$$|S(p, q)| = d(s, p) + d(p, \bar{p}) = d(s, p) + |pq| + \delta(p, q)$$

$$|E(p,q)| = d(e,q) + d(q,\bar{q}) = d(e,q) + |pq| + \delta(p,q)$$

The following lemma characterizes which of the paths $U(p,q)$, $S(p,q)$, and $E(p,q)$ determine the diameter of $P + pq$. Notice that these cases overlap, for instance, $E(p,q)$ and $S(p,q)$ are both continuous diametral when $d(s,p) = d(e,q) \leq \delta(p,q)$.

▶ **Lemma 2.2.** *Let $pq$ be a shortcut for a path $P$. Let $x = d(s,p)$, $y = d(e,q)$, and $z = \delta(p,q)$.*
- *The path $U(p,q)$ is continuous diametral if and only if $z = \min(x,y,z)$.*
- *The path $S(p,q)$ is continuous diametral if and only if $y = \min(x,y,z)$.*
- *The path $E(p,q)$ is continuous diametral if and only if $x = \min(x,y,z)$.*

**Proof.** The claim follows, since the following relations are preserved for $\sim \in \{<,=,>\}$.

$$|U| \sim |S| \iff d(s,p) + |pq| + d(q,e) \sim d(s,p) + |pq| + \delta(p,q) \iff d(e,q) \sim \delta(p,q)$$
$$|U| \sim |E| \iff d(s,p) + |pq| + d(q,e) \sim d(e,q) + |pq| + \delta(p,q) \iff d(s,p) \sim \delta(p,q)$$
$$|S| \sim |E| \iff d(s,p) + |pq| + \delta(p,q) \sim d(e,q) + |pq| + \delta(p,q) \iff d(s,p) \sim d(e,q)$$

For instance, we have $|U(p,q)| \geq |S(p,q)|$ and $|U(p,q)| \geq |E(p,q)|$ if and only if $d(e,q) \geq \delta(p,q)$ and $d(s,p) \geq \delta(p,q)$, i.e., $z = \delta(p,q) = \min(d(s,p), d(e,q), \delta(p,q)) = \min(x,y,z)$. ◀

▶ **Lemma 2.3.** *For every path $P$, there is an optimal shortcut $pq$ such that $S(p,q)$ and $E(p,q)$ are continuous diametral paths in $P + pq$, i.e., $\mathrm{diam}(P + pq) = |S(p,q)| = |E(p,q)|$.*

**Proof.** Suppose we have a shortcut $pq$ for a path $P$ such that $U(p,q)$ is continuous diametral in $P + pq$. In the notation of Lemma 2.2, this means $z \leq x$ and $z \leq y$.

By the triangle inequality, the length of $U(p,q)$—and, thus, the continuous diameter—decreases or remains the same as we move $p$ closer to $s$ (decreasing $x$) or $q$ closer to $e$ (decreasing $y$). Moreover, decreasing $x$ or $y$ increases $z$. Therefore, we can move the shortcut closer to $s$ and to $e$ until we have $x = z$ or $y = z$, i.e., $\min(x,y,z) = \min(x,y)$ while maintaining or decreasing the continuous diameter. According to Lemma 2.2, $S(p',q')$ or $E(p',q')$ are continuous diametral in $P + p'q'$ for the resulting shortcut $p'q'$. Consequently, there is an optimal shortcut $p^*q^*$ where $S(p^*,q^*)$ or $E(p^*,q^*)$ are continuous diametral.

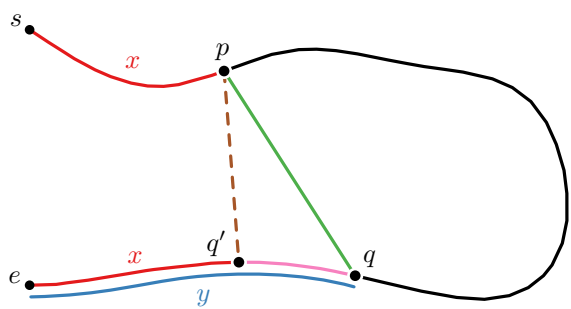

**Figure 6** Moving a shortcut to balance the distances to the endpoints of the path.

Now suppose $E(p,q)$ is a continuous diametral path and $S(p,q)$ is not, i.e., $d(s,p) < d(e,q)$, according to Lemma 2.2. Let $q' \in P$ be the point with $d(s,p) = d(e,q')$, as illustrated in Figure 6. By the triangle inequality, we have $|pq| + d(q,q') \geq |pq'|$ and, thus,

$$|E(p,q)| = d(e,q) + \frac{|pq| + d(p,q)}{2} = d(e,q') + d(q,q') + \frac{|pq| + d(p,q)}{2}$$

$$= d(e, q') + \frac{|pq| + d(q, q') + d(p, q) + d(q, q')}{2}$$
$$\geq d(e, q') + \frac{|pq'| + d(p, q')}{2} = |E(p, q')| \ .$$

We have $|E(p, q')| = |S(p, q')|$ by construction and we have $|E(p, q')| \geq |U(p, q')|$ because of the following. Since $E(p, q)$ was continuous diametral, we have $|E(p, q)| \geq |U(p, q)|$ and $d(s, p) \leq \delta(p, q)$ by Lemma 2.2. This implies $\delta(p, q') \geq d(s, p)$, because

$$\delta(p, q) = \frac{d(p, q) - |pq|}{2} = \frac{d(p, q) + d(q, q') - |pq| - d(q, q')}{2}$$
$$= \frac{d(p, q') - |pq| - d(q, q')}{2}$$
$$\leq \frac{d(p, q') - |pq'|}{2} = \delta(p, q') \ .$$

Using $\delta(p, q') \geq d(s, p)$ and $d(s, p) = d(e, q')$, we obtain $|E(p, q')| \geq |U(p, q')|$, as

$$|E(p, q')| = d(e, q') + \delta(p, q') + |pq'| \geq d(s, p) + d(e, q') + |pq'| = |U(p, q')| \ .$$

In summary, we have $\operatorname{diam}(P + pq) = |E(p, q)| \geq |E(p, q')| = |S(p, q')| = \operatorname{diam}(P + pq')$, i.e., there is an optimal shortcut $p^*q^*$ where $S(p^*, q^*)$ and $E(p^*, q^*)$ are continuous diametral. ◀

According to Lemmas 2.2 and 2.3, we can restrict our search for an optimal shortcut to those shortcuts satisfying $d(s, p) = d(e, q) \leq \delta(p, q)$. For $x \in [0, |P|/2]$, let $p(x)$ and $q(x)$ be the points on $P$ such that $x = d(s, p(x))$ and $x = d(e, q(x))$, and let $D(x) = |p(x)q(x)|$. Notice that $d(p(x), q(x)) = |P| - 2x$. Using this notation, we phrase our problem as

$$\text{minimize } x + \frac{d(p(x), q(x)) + |p(x)q(x)|}{2} = x + \frac{|P| - 2x + |p(x)q(x)|}{2} = \frac{|P| + D(x)}{2}$$
$$\text{such that } x \leq \delta(p(x), q(x)) = \frac{d(p(x), q(x)) - |p(x)q(x)|}{2} = \frac{|P| - 2x - D(x)}{2} \ ,$$

which simplifies to minimizing $D(x)$ such that $4x + D(x) \leq |P|$.

▶ **Lemma 2.4.** *The function $B(x) = 4x + D(x)$ is strictly increasing on $[0, |P|/2]$.*

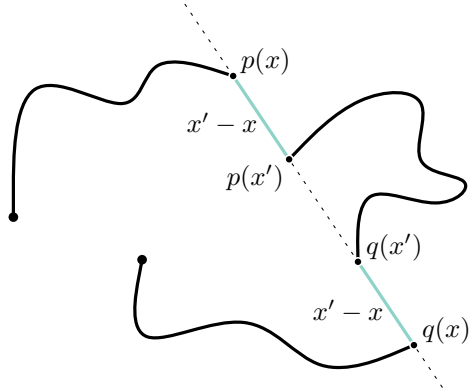

**Figure 7** Moving each endpoint of the shortcut by a geodesic distance of $x' - x$ along the path decreases their Euclidean distance by at most $2(x' - x)$, attained when moving towards each other.

**Proof.** Let $x, x' \in [0, |P|/2]$ with $x < x'$. As illustrated in Figure 7, the length of the shortcut decreases by at most $2(x' - x)$ from $p(x)q(x)$ to $p(x')q(x')$, because each endpoint moves a distance of $x' - x$. Thus, we have $D(x') - D(x) \geq -2(x' - x)$ and the claim follows, since

$$\begin{aligned} B(x') - B(x) &= 4x' + D(x') - 4x - D(x) \\ &= 4(x' - x) + D(x') - D(x) \\ &\geq 4(x' - x) - 2(x' - x) \\ &= 2(x' - x) > 0 \ . \end{aligned}$$
◀

Summarizing the above, the following theorem describes an optimal shortcut.

▶ **Theorem 2.5.** *Let $P$ be a path and let $b$ be the unique value in $[0, |P|/2]$ with $B(b) = |P|$. Suppose $D$ has a global minimum in the interval $[0, b]$ at $x^*$, i.e., $D(x^*) = \min_{x \in [0,b]} D(x)$. Then the shortcut $p(x^*)q(x^*)$ achieves the minimum continuous diameter for $P$.*

**Proof.** As argued above, an optimal shortcut is attained at the minimum of $D(x)$ among those $x \in [0, |P|/2]$ satisfying $4x + D(x) \leq |P|$. The claim follows, because these values of $x$ form an interval $[0, b]$, where $b$ is the unique value in $[0, |P|/2]$ with $B(b) = |P|$, since $B(0) = 4 \cdot 0 + |p(0)q(0)| = |se| \leq |P|$ and since $B$ is strictly increasing due to Lemma 2.4. ◀

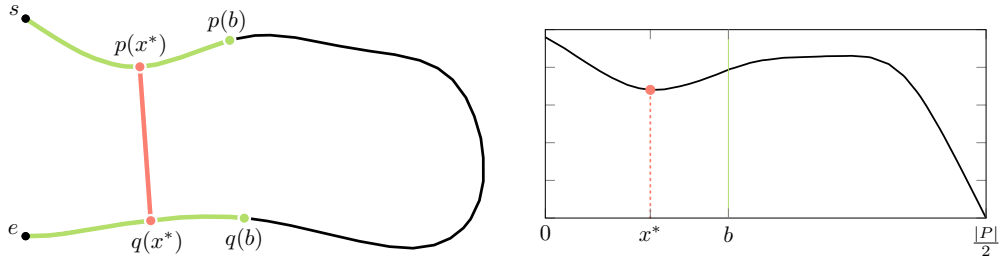

**(a)** A path $P$ with its optimal shortcut $p(x^*)q(x^*)$. **(b)** A plot of $D(x) = |p(x)q(x)|$.

**Figure 8** The optimal shortcut for the path in (a) with the function $D(x)$ plotted in (b).

▶ **Lemma 2.6.** *Let $P$ be a path with $n$ vertices. Then $D^2(x)$ is a continuous function whose graph consists of at most $n$ parabolic arcs or line segments.*

**Proof.** Let $x_1, x_2 \in [0, |P|/2]$ be such that $v_1 = p(x_1)$ and $v_2 = p(x_2)$ lie on the same edge of $P$ and $u_1 = q(x_1)$ and $u_2 = q(x_2)$ lie on the same edge of $P$. With $\lambda(x) = \frac{x - x_1}{x_2 - x_1}$, we have

$$p(x) = (1 - \lambda(x)) \cdot v_1 + \lambda(x) \cdot v_2 \text{ and } q(x) = (1 - \lambda(x)) \cdot u_1 + \lambda(x) \cdot u_2$$

for all $x \in [x_1, x_2]$. Let $\langle u, v \rangle$ denote the scalar product of two vectors $u$ and $v$. Expanding $D^2(x) = \|p(x) - q(x)\|^2$ with the above representations of $p(x)$ and $q(x)$ yields

$$\begin{aligned} D^2(x) &= \|(1 - \lambda(x)) \cdot (v_1 - u_1) + \lambda(x) \cdot (v_2 - u_2)\|^2 \\ &= (1 - \lambda(x))^2 \cdot |v_1 u_1|^2 + 2(1 - \lambda(x))\lambda(x) \cdot \langle v_1 - u_1, v_2 - u_2 \rangle + \lambda(x)^2 \cdot |v_2 u_2|^2 \ . \end{aligned}$$

This means that $D^2(x)$ is a convex parabola with its apex at

$$a = (x_2 - x_1) \cdot \frac{\langle v_1 - u_1, v_1 - v_2 + u_2 - u_1 \rangle}{\|v_1 - v_2 + u_2 - u_1\|^2} + x_1$$

when $v_2 - v_1 \neq u_2 - u_1$ and $D^2(x)$ has constant value $|v_1 u_1|^2$ when $v_2 - v_1 = u_2 - u_1$. The latter case applies when $p(x)$ and $q(x)$ travel in the same direction maintaining their distance.

When $v_2 - v_1 \neq u_2 - u_1$ and $x_1 \leq a \leq x_2$, the minimum of $D^2(x)$ on $[x_1, x_2]$ is

$$D^2(a) = \frac{|v_1 u_1|^2 |v_2 u_2|^2 - \langle v_1 - u_1,\, v_2 - u_2 \rangle^2}{\|v_1 - v_2 + u_2 - u_1\|^2} \; ,$$

and the minimum of $D^2(x)$ on $[x_1, x_2]$ is $\min(|u_1 v_1|^2, |u_2 v_2|^2)$, otherwise.

We subdivide $[0, |P|/2]$ at those values of $x$ where $p(x)$ or $q(x)$ coincides with a vertex. This yields at most $n$ intervals on each of which $D^2$ is a parabolic arc or a horizontal segment (i.e., a degenerate parabolic arc), since $p(x)$ and $q(x)$ only switch edges at a vertex. ◀

▶ **Corollary 2.7.** *Given a path $P$ with $n$ vertices, we can compute a shortcut for $P$ achieving the minimal continuous diameter in $O(n)$ time.*

**Proof.** Let $x_{\pi(1)} \leq x_{\pi(2)} \leq \cdots \leq x_{\pi(n)}$ be the values in $[0, |P|/2]$ where $p(x_{\pi(i)})$ or $q(x_{\pi(i)})$ coincides with the $i$-th vertex of $P$ for each $i = 1, 2, \ldots, n$.

We compute the minimum of the parabolic arc of $D^2$ on each interval $[x_{\pi(i)}, x_{\pi(i+1)}]$ for $i = 1, 2, \ldots, n$ until we arrive at $k$ with $B(x_{\pi(k)}) < |P|$ and $B(x_{\pi(k+1)}) \geq |P|$. We then compute $b$ by solving the quadratic equation $D^2(b) = (|P| - 4b)^2$ and, finally, compute the minimum of $D^2(x)$ on $[x_{\pi(k)}, b]$. The lowest minima of the encountered parabolic arcs is the global minimum of $D$ on $[0, b]$, which reveals the position of an optimal shortcut according to Theorem 2.5. Altogether, the running time is $\Theta(k+1) = O(n)$, since we obtain $x_{\pi(1)}, x_{\pi(2)}, \ldots, x_{\pi(k+1)}$ by merging the vertices by their distances from $s$ or from $e$. ◀

▶ **Remark.** Our result on the location of an optimal shortcut from Theorem 2.5 also holds for rectifiable curves in the plane. However, obtaining an optimal shortcut for such curves depends on our ability to calculate $b$ and a global minima of $D(x)$ in the interval $[0, b]$.

## 3 Shortcuts for Cycles

Consider a polygonal cycle $C$ in the plane; $C$ may have crossings but cannot lie on a line. For any two points $p$ and $q$ along $C$ that may be vertices or points along edges of $C$, let $d_{\text{ccw}}(p, q)$ and $d_{\text{cw}}(p, q)$ be their counter-clockwise and clockwise distance along $C$, respectively. Let $d(p, q) = \min(d_{\text{ccw}}(p, q), d_{\text{cw}}(p, q))$ denote the geodesic distance between $p$ and $q$ along $C$. We call the line segment $pq$ a shortcut for $C$ when $|pq| < d(p, q)$. We seek to minimize the continuous diameter by augmenting $C$ with shortcuts.

▶ **Lemma 3.1.** *Adding a single shortcut $pq$ to a polygonal cycle $C$ never decreases the continuous diameter, i.e., $\mathrm{diam}(C) \leq \mathrm{diam}(C + pq)$ for all $p, q \in C$.*

**Proof.** Consider any shortcut $pq$ to a cycle $C$. Let $C_{\text{ccw}}$ be the cycle consisting of $pq$ and the counter-clockwise path from $p$ to $q$ along $C$, as illustrated in Figure 9. Let $\bar{p}_{\text{ccw}}$ and $\bar{q}_{\text{ccw}}$ be the farthest points from $p$ and from $q$ on $C_{\text{ccw}}$, respectively. Since $\bar{p}_{\text{ccw}}$ and $\bar{q}_{\text{ccw}}$ are antipodal from $p$ and $q$ in $C_{\text{ccw}}$, we have $d(\bar{q}_{\text{ccw}}, \bar{p}_{\text{ccw}}) = |pq|$ and $d(\bar{p}_{\text{ccw}}, q) = d(p, \bar{q}_{\text{ccw}})$.

Consider a point $g$ along the clockwise path from $\bar{p}_{\text{ccw}}$ to $\bar{q}_{\text{ccw}}$ and let $\bar{g} \in C$ be the farthest-point form $g$ with respect to $C$. We claim that $d_C(g, \bar{g}) = d_{C+pq}(g, \bar{g})$.

Assume that there is a shortest path from $g$ to $\bar{g}$ in $C + pq$ that contains $pq$ and is shorter than the path from $g$ to $\bar{g}$ along $C$. Suppose this path reaches $q$ before $p$. Since $p$ lies on at least one shortest path from $g$ to $\bar{g}$ and since $\bar{p}_{\text{ccw}}$ lies on the path from $g$ to $q$, we have

$$d_{C+pq}(g, \bar{g}) = d_C(g, q) + |pq| + d_C(p, \bar{g}) = d_C(g, \bar{p}_{\text{ccw}}) + d_C(\bar{p}_{\text{ccw}}, q) + |pq| + d_C(p, \bar{g}) \; ,$$

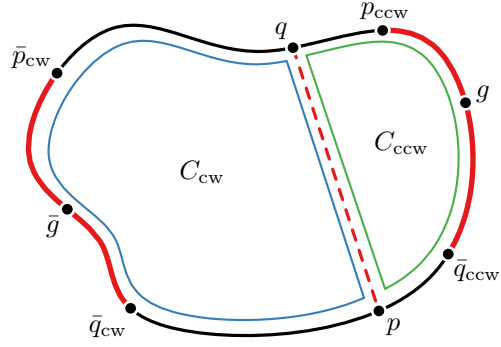

**Figure 9** The unaffected regions (solid red) for a shortcut $pq$ (dashed red) to a cycle. The point $\bar{x}_y$ denotes the farthest points from $x$ along the cycle $C_y$ for $x \in \{p,q\}$ and $y \in \{\text{cw}, \text{ccw}\}$. Any point $g$ along the clockwise path from $\bar{p}_{\text{ccw}}$ to $\bar{q}_{\text{ccw}}$ has their farthest point $\bar{g}$ on the clockwise path from $\bar{q}_{\text{cw}}$ to $\bar{p}_{\text{cw}}$ and vice versa. The distance between $g$ and $\bar{g}$ is unaffected by the addition of $pq$ to $C$.

and we can express the distance between $g$ and $\bar{g}$ along $C$ as

$$d_C(g, \bar{g}) = d_C(g,p) + d_C(p, \bar{g}) = d_C(g, \bar{q}_{\text{ccw}}) + d_C(\bar{q}_{\text{ccw}}, p) + d_C(p, \bar{g}) \ .$$

Together with $d_C(\bar{p}_{\text{ccw}}, q) = d_C(p, \bar{q}_{\text{ccw}})$ and $d_C(\bar{q}_{\text{ccw}}, g) + d_C(g, \bar{p}_{\text{ccw}}) = |pq|$, our assumption $d_C(g, \bar{g}) < d_{C+pq}(g, \bar{g})$ implies the following.

$$d_C(g, \bar{p}_{\text{ccw}}) + d_C(\bar{p}_{\text{ccw}}, q) + |pq| + d_C(p, \bar{g})$$
$$< d_C(g, \bar{q}_{\text{ccw}}) + d_C(\bar{q}_{\text{ccw}}, p) + d_C(p, \bar{g}) = |pq| - d_C(g, \bar{p}_{\text{ccw}}) + d_C(\bar{p}_{\text{ccw}}, q) + d_C(p, \bar{g})$$

This implies the contradiction $d_C(g, \bar{p}_{\text{ccw}}) < 0$. Hence, no shortest path from $g$ to $\bar{g}$ in $C + pq$ contains $pq$ and, thus, $\text{diam}(C) = d_C(g, \bar{g}) = d_{C+pq}(g, \bar{g}) \leq \text{diam}(C + pq)$. ◀

According to Lemma 3.1, some points preserve their farthest distance in $C$ when adding a single shortcut $pq$ to $C$. The points that are unaffected by $pq$ in this sense form the *unaffected region* of $pq$ that consists of the counter-clockwise path from $\bar{q}_{\text{ccw}}$ to $\bar{p}_{\text{ccw}}$ and the clockwise path from $\bar{q}_{\text{cw}}$ to $\bar{p}_{\text{cw}}$, as illustrated in Figure 9. Conversely, every point on $C$ outside of the unaffected region uses $pq$ as a shortcut to their farthest point in $C + pq$.

Consequently, we have to add at least two shortcuts $pq$ and $rs$ in order to decrease the continuous diameter of the augmented cycle $C + pq + rs$. We call a pair of shortcuts $pq$ and $rs$ *useful* when $\text{diam}(C) > \text{diam}(C + rs + pq)$, and we call $pq$ and $rs$ *useless*, otherwise. A pair of shortcuts $pq$ and $rs$ is useful if and only if their unaffected regions are disjoint.

We call a polygonal cycle $C$ *degenerate* when it consists of two congruent line segments of length $|C|/2$. No number of shortcuts can decrease the diameter of a degenerate cycle, since the endpoints of its line segment will always remain at distance $\text{diam}(C) = |C|/2$.

▶ **Theorem 3.2.** *For every non-degenerate cycle $C$, there exists a pair of shortcuts $pq$ and $rs$ that decrease the continuous diameter, i.e., $\text{diam}(C) > \text{diam}(C + pq + rs)$.*

**Proof.** Let $C$ be a cycle. Suppose there exist three points $p$, $q$, and $s$ on $C$ with $d(p,q) = d(q,s) = |C|/4$ such that $pq$ and $qs$ are shortcuts, i.e., $|pq| < d(p,q)$ and $|qs| < d(q,s)$.

We argue that $pq$ and $qs$ are useful. Let $\bar{q}$ denote the farthest point from $q$ on $C$. As illustrated in Figure 10, the unaffected region of $pq$ is confined to the interior of the clockwise paths from $q$ to $p$ and from $\bar{q}$ to $s$, and the unaffected region of $qs$ is confined to the interior of the clockwise paths from $s$ to $q$ and from $p$ to $\bar{q}$. Therefore, the unaffected regions of $pq$ and $qs$ are disjoint, i.e., $pq$ and $qs$ are useful shortcuts, i.e., $\mathrm{diam}(C) > \mathrm{diam}(C + pq + qs)$.

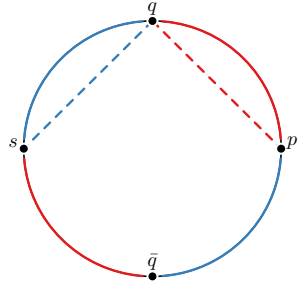

**Figure 10** The unaffected regions of two touching shortcuts $pq$ and $qs$.

Suppose, on the other hand, that for every three points $p$, $q$, and $s$ on $C$ with $d(p,q) = d(q,s) = |C|/4$ at least one of $pq$ and $qs$ is not a shortcut, i.e., $|pq| = d(p,q)$ or $|qs| = d(q,s)$. We argue that $C$ is degenerate by showing that $C$ contains a line segment of length $|C|/2$.

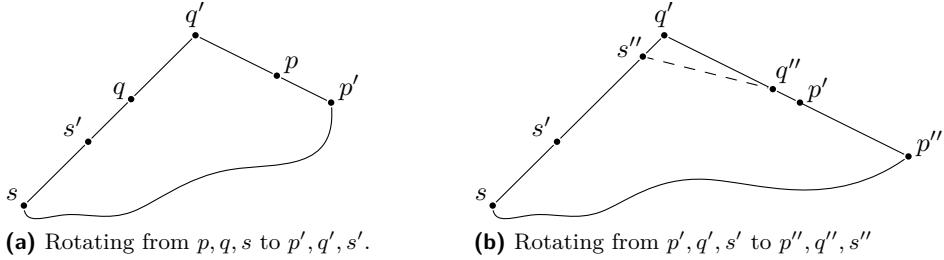

(a) Rotating from $p, q, s$ to $p', q', s'$.   (b) Rotating from $p', q', s'$ to $p'', q'', s''$.

**Figure 11** Illustration of the discussion regarding the shape of a cycle where one of the line segments $sq$ or $qp$ is not a shortcut for every tripel of points with $d(s,q) = d(q,p) = |C|/4$.

Assume, without loss of generality, that $qs$ is not a shortcut, i.e., $C$ contains the line segment $qs$. We move $p$, $q$, and $s$ clockwise along $C$ while maintaining $d(p,q) = d(q,s) = |C|/4$ until we arrive at the first positions $p'$, $q'$ and $s'$ where $q's'$ ceases to be a shortcut, i.e., $C$ contains the line segment $sq'$ and the line segment $q'p'$, as illustrated in Figure 11a. The points $p'$, $q'$ and $s'$ exist, since $0 \leq |sq'| \leq |C|/2$, because the cycle $C$ cannot contain a line segment that is longer than $|C|/2$. If $|sq'| = |C|/2$, $C$ is degenerate. Suppose $|sq'| < |C|/2$.

We move the three points again by a distance of $|C|/4 - \epsilon$ for some $\epsilon$ with $0 < \epsilon < |C|/4$. Let $p''$, $q''$, and $s''$ be the resulting points. As illustrated in Figure 11b, $s''$ lies on $s'q'$ at distance $\epsilon$ from $q'$, and $q''$ lies on $q'p'$ at distance $\epsilon$ from $p'$. Since $d(s'',q'') = d(q'',p'') = |C|/4$, one of $s''q''$ or $q''p''$ is not a shortcut. However, $s''q''$ must be a shortcut, since otherwise $q'$ would be contained in the line segment $s''q''$ contradicting the choice of $q'$ as the first point where $q'p'$ is not a shortcut. Therefore, $q''p''$ is not a shortcut and $C$ contains the line segment $q'p''$ whose length is $|q'p''| = |q'p'| + |p'p''| = |C|/2 - \epsilon$. Since the above holds for arbitrary small values of $\epsilon$, the cycle $C$ contains a line segment of length $|C|/2$. ◀

## 3.1   Alternating vs. Consecutive

When placing two shortcuts $pq$ and $rs$ on a cycle $C$, we distinguish whether their endpoints appear in alternating order or in consecutive order along the cycle, as illustrated in Figure 12. We show that there is always an optimal pair of shortcuts in the alternating configuration.

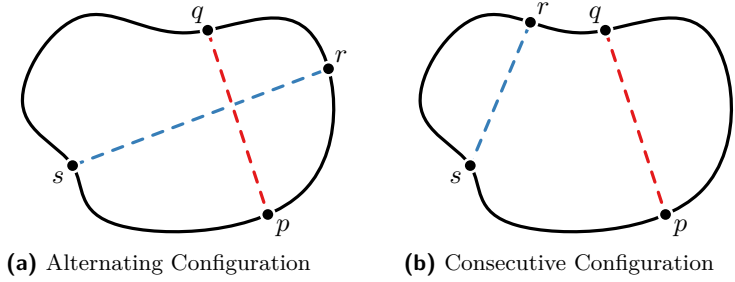

**(a)** Alternating Configuration    **(b)** Consecutive Configuration

**Figure 12** The two cases for adding two shortcuts $pq$ and $rs$ to a cycle $C$. The endpoints of the shortcuts appear in alternating cyclic order $p$, $r$, $q$, and $s$, as shown in (a), or in consecutive cyclic order $p$, $q$, $r$, and $s$, as shown in (b). The two cases overlap when $q$ coincides with $r$.

To establish our claim we study the cycles created by the insertion of the shortcuts. We call a cycle in $C + pq + rs$ *diametral* when it contains a diametral pair. Each configuration has five candidates for diametral cycles: two that use both shortcuts, two that use one of the shortcuts, and one ($C$) that does not use any shortcut. Figures 13 and 14 illustrate the candidates for diametral cycles in each configuration, except for the cycle $C$ itself. To help distinguish the cycles using one shortcut, we color $pq$ red and $rs$ blue and we refer to the longer cycle in $C + pq + rs$ using the red shortcut $pq$ as the *red split* and we refer to the longer cycle using the blue shortcut $rs$ as the *blue split*. If $C$ happens to be diametral in $C + pq + rs$, then our pair of shortcuts is useless.

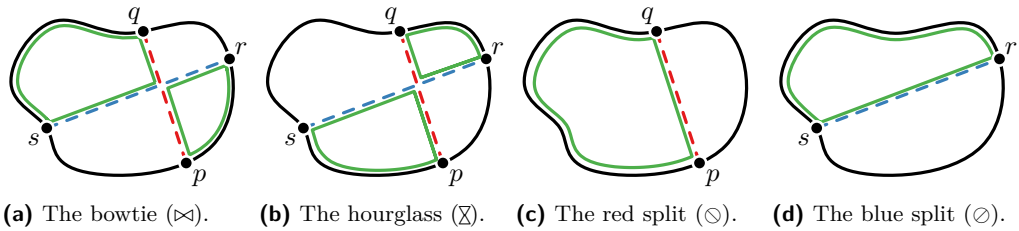

**(a)** The bowtie (⋈).   **(b)** The hourglass (⧖).   **(c)** The red split (⊘).   **(d)** The blue split (⊘).

**Figure 13** The candidate diametral cycles, except $C$, for shortcuts in the alternating configuration.

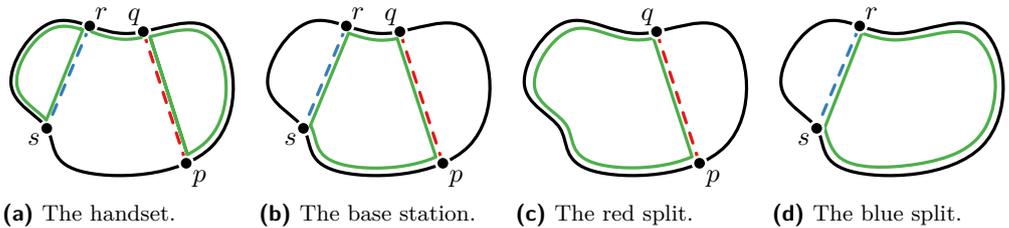

**(a)** The handset.   **(b)** The base station.   **(c)** The red split.   **(d)** The blue split.

**Figure 14** The candidate diametral cycles, except $C$, for shortcuts in the consecutive configuration, depicted for $d(q, r) \leq d(s, p)$. Even though the handset (a) is no simple cycle, it might still contain a diametral pair. Observe that the base station (b) is only listed for the sake of completeness: by the triangle inequality, this cycle is never longer than the split cycles and, therefore, never diametral.

For the following, let the points $\bar{x}_y$ with $x \in \{p, q, r, s\}$ and $y \in \{\text{cw}, \text{ccw}\}$ be defined as in Figure 9, e.g., let $\bar{r}_{\text{cw}}$ be the farthest point from $r$ on the cycle $C_{\text{cw}}(r, s)$ that consists of $rs$ and the clockwise path from $r$ to $s$ along $C$, and let $\bar{q}_{\text{ccw}}$ be the farthest point from $q$ on the cycle $C_{\text{ccw}}(p, q)$ that consists of $pq$ and the counter-clockwise path from $p$ to $q$.

▶ **Lemma 3.3.** *Two shortcuts $pq$ and $rs$ in alternating configuration are useful if and only if $|pq| + |rs| < d_{ccw}(r, q) + d_{ccw}(s, p)$ and $|pq| + |rs| < d_{ccw}(p, r) + d_{ccw}(q, s)$.*

**Proof.** Suppose $pq$ and $rs$ are useless, i.e., the unaffected regions of $pq$ and $rs$ overlap. In the alternating configuration, this overlap occurs along the bowtie or along the hourglass. Since these cases are symmetric, we consider only the former in the following.

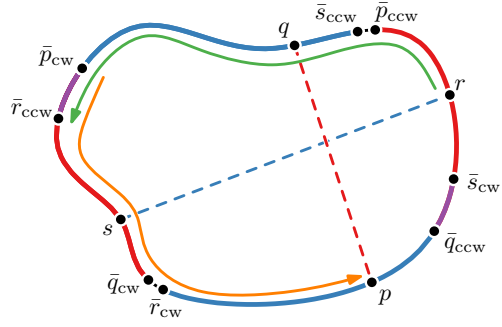

**Figure 15** A pair of useless shortcuts whose unaffected regions have an overlap (purple) along the bowtie, i.e., the points $s$, $\bar{r}_{\text{ccw}}$, $\bar{p}_{\text{cw}}$, and $q$ appear clockwise in this order along the cycle.

An overlap on the bowtie manifests along the clockwise path from $\bar{r}_{\text{ccw}}$ to $\bar{p}_{\text{cw}}$ with a mirrored overlap along the clockwise path from $\bar{s}_{\text{cw}}$ to $\bar{q}_{\text{ccw}}$, as illustrated in Figure 15. This means the sum of the lengths of the counter-clockwise paths from $r$ to $\bar{r}_{\text{ccw}}$ and from $\bar{p}_{\text{cw}}$ to $p$ is at least the length of the counter-clockwise path from $r$ to $p$, i.e., $d_{\text{ccw}}(\bar{p}_{\text{cw}}, p) + d_{\text{ccw}}(r, \bar{r}_{\text{ccw}}) \geq d_{\text{ccw}}(r, p)$. This is equivalent to $|pq| + |rs| \geq d_{\text{ccw}}(r, q) + d_{\text{ccw}}(s, p)$, since

$$d_{\text{ccw}}(q, p) + |pq| + d_{\text{ccw}}(r, s) + |rs| = 2d_{\text{ccw}}(\bar{p}_{\text{cw}}, p) + 2d_{\text{ccw}}(r, \bar{r}_{\text{ccw}}) \geq 2d_{\text{ccw}}(r, p)$$
$$\iff |pq| + |rs| \geq \underbrace{d_{\text{ccw}}(r, p) - d_{\text{ccw}}(q, p)}_{=d_{\text{ccw}}(r,q)} + \underbrace{d_{\text{ccw}}(r, p) - d_{\text{ccw}}(r, s)}_{=d_{\text{ccw}}(s,p)} .$$

Analogously, we derive that $|pq| + |rs| \geq d_{\text{ccw}}(p, r) + d_{\text{ccw}}(q, s)$ holds if and only if there is an overlap along the hourglass. Consequently, the shortcuts $pq$ and $rs$ are useful if and only if $|pq| + |rs| < d_{\text{ccw}}(r, q) + d_{\text{ccw}}(s, p)$ and $|pq| + |rs| < d_{\text{ccw}}(p, r) + d_{\text{ccw}}(q, s)$. ◀

▶ **Lemma 3.4.** *Consider two consecutive shortcuts $pq$ and $rs$ with $d_{ccw}(q, r) \leq d_{ccw}(s, p)$. Then $pq$ and $rs$ are useful if and only if $|pq| + |rs| < d_{ccw}(s, p) - d_{ccw}(q, r)$.*

**Proof.** Suppose $pq$ and $rs$ are useless, i.e., the unaffected regions of $pq$ and $rs$ overlap. In the consecutive configuration with $d_{\text{ccw}}(q, r) \leq d_{\text{ccw}}(s, p)$, this overlap occurs on the handset and manifests along the clockwise path from $\bar{r}_{\text{ccw}}$ to $\bar{p}_{\text{cw}}$ with a mirrored overlap along the clockwise path from $\bar{s}_{\text{cw}}$ to $\bar{q}_{\text{ccw}}$. An overlap along a clockwise path from $\bar{p}_{\text{ccw}}$ to $\bar{s}_{\text{ccw}}$ is impossible, since otherwise $\bar{p}_{\text{ccw}}$ would lie on $C_{\text{cw}}(p, q)$ or $\bar{s}_{\text{ccw}}$ would lie on $C_{\text{cw}}(r, s)$, contradicting their definitions. We argue below that an overlap along a clockwise path from $\bar{q}_{\text{cw}}$ to $\bar{r}_{\text{cw}}$ is impossible, as well, after discussing an overlap along the handset.

Suppose we have an overlap on the handset as illustrated in Figure 16. This occurs when the sum of the lengths of the counter-clockwise paths from $r$ to $\bar{r}_{\text{ccw}}$ and from $\bar{p}_{\text{cw}}$ to $p$ is at least the length of the counter-clockwise path from $r$ to $p$, i.e., $d_{\text{ccw}}(\bar{p}_{\text{cw}}, p) + d_{\text{ccw}}(r, \bar{r}_{\text{ccw}}) \geq d_{\text{ccw}}(r, p)$. This is equivalent to $|pq| + |rs| \geq d_{\text{ccw}}(s, p) - d_{\text{ccw}}(q, r)$, since

$$d_{\text{ccw}}(q, p) + |pq| + d_{\text{ccw}}(r, s) + |rs| = 2d_{\text{ccw}}(\bar{p}_{\text{cw}}, p) + 2d_{\text{ccw}}(r, \bar{r}_{\text{ccw}}) \geq 2d_{\text{ccw}}(r, p)$$

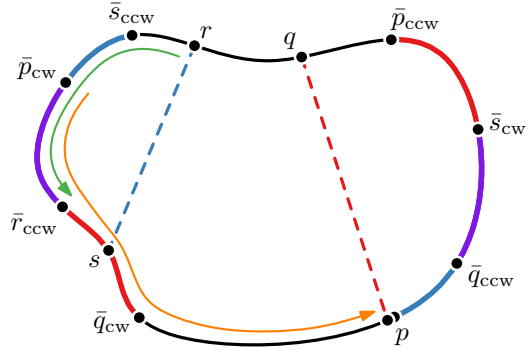

**Figure 16** A pair of useless shortcuts whose unaffected regions overlap on the handset.

$$\iff |pq| + |rs| \geq \underbrace{d_{\text{ccw}}(r,p) - d_{\text{ccw}}(r,s)}_{=d_{\text{ccw}}(s,p)} + \underbrace{d_{\text{ccw}}(r,p) - d_{\text{ccw}}(q,p)}_{=-d_{\text{ccw}}(q,r)} \ .$$

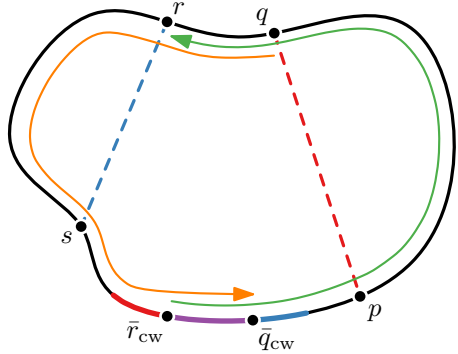

**Figure 17** An impossible overlap for shortcuts in the consecutive configuration.

Suppose we have an overlap on the base station along a clockwise path from $\bar{q}_{\text{cw}}$ to $\bar{r}_{\text{cw}}$, as illustrated in Figure 17. This occurs when the sum of the lengths of the clockwise paths from $r$ to $\bar{r}_{\text{cw}}$ and from $\bar{q}_{\text{cw}}$ to $q$ is at least the length of the cycle $C$, i.e., $d_{\text{ccw}}(\bar{r}_{\text{cw}}, r) + d_{\text{ccw}}(q, \bar{q}_{\text{cw}}) \geq |C| + d_{\text{ccw}}(q,r) \geq |C|$. This is equivalent to $|pq| + |rs| \geq d(p,q) + d(r,s)$, since

$$d_{\text{ccw}}(q,p) + |pq| + d_{\text{ccw}}(s,r) + |rs| = 2d_{\text{ccw}}(q,\bar{q}_{\text{cw}}) + 2d_{\text{ccw}}(\bar{r}_{\text{cw}},r) \geq 2|C|$$
$$\iff |pq| + |rs| \geq \underbrace{|C| - d_{\text{ccw}}(q,p)}_{=d_{\text{ccw}}(p,q) \geq d(p,q)} + \underbrace{|C| - d_{\text{ccw}}(s,r)}_{=d_{\text{ccw}}(r,s) \geq d(r,s)} \ .$$

However, the inequality $|pq| + |rs| \geq d(p,q) + d(r,s)$ implies $|pq| \geq d(p,q)$ or $|rs| \geq d(r,s)$, i.e., one of $pq$ or $rs$ was not a shortcut to begin with contradicting their choice. Thus, the unaffected regions in the consecutive case can only overlap along the handset.    ◀

▶ **Theorem 3.5.** *Let $pq$ and $rs$ be a pair of shortcuts for a cycle $C$ in consecutive configuration. There exists a pair $p'q'$ and $r's'$ of shortcuts in the alternating configuration that are at least as good as $pq$ and $rs$, i.e., $\text{diam}(C + p'q' + r's') \leq \text{diam}(C + pq + rs)$.*

**Proof.** Suppose $pq$ and $rs$ are useful shortcuts in the consecutive configuration. Assume, without loss of generality, $d_{\text{ccw}}(q,r) \leq d_{\text{ccw}}(s,p)$ and $d_{\text{ccw}}(p,q) \leq d_{\text{ccw}}(r,s)$.

We consider the shortcuts $p'q' = pr$ and $r's' = rs$, which are illustrated in Figure 18 and lie in the intersection of the alternating and consecutive case. We argue that $pr$ and $rs$ are useful shortcuts and that each candidate diametral cycle in $C + pq + rs$ has a one-to-one correspondence to a candidate diametral cycle in $C + pr + rs$ of smaller or equal length.

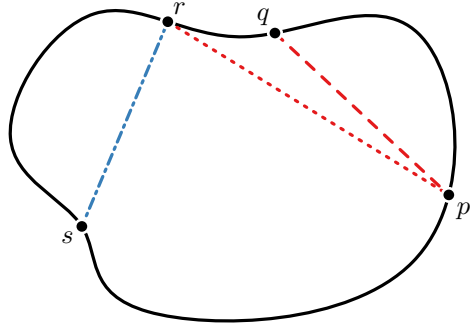

▮ **Figure 18** Replacing two shortcuts $pq$ and $rs$ in consecutive configuration with two shortcuts $p'q' = pr$ and $r's' = rs$ that are in both consecutive and alternating configuration.

The line segment $pr$ is a shortcut, because otherwise $|pr| = d(p,r)$, i.e., the shortest path from $p$ to $r$ along $C$ is a line segment containing $q$. This would imply $|pq| = d(p,q)$ contradicting $pq$ being a shortcut. We have $|pr| < d(p,r)$ and $|rs| < d(r,s)$, since $pr$ and $rs$ are shortcuts, and we have $|pq| + |rs| + d_{\text{ccw}}(q,r) < d_{\text{ccw}}(s,p)$, since $pq$ and $rs$ are useful, by Lemma 3.4. According to Lemma 3.3, $pr$ and $rs$ are also useful, since

$$|pr| + |rs| < d(p,r) + d(r,s) \leq d_{\text{ccw}}(p,r) + d_{\text{ccw}}(r,s) \ ,$$
and $|pr| + |rs| \leq |pq| + d(q,r) + |rs| < d_{\text{ccw}}(s,p) \leq d_{\text{ccw}}(s,p) + d_{\text{ccw}}(r,r) \ .$

The bowtie in $C + pr + rs$ is at most as long as the handset in $C + pq + rs$ and the hourglass in $C + pr + rs$ is at most as long as the base station in $C + pq + rs$. The blue split cycle remains unchanged, as $r's' = rs$. Below, we proof that the red split cycle in $C + pr + rs$ is at most as long as the red split cycle in $C + pq + rs$.

First, the red split cycle remains on the same side, viz., clockwise, of the red shortcut when replacing $pq$ with $pr$. Our premises $d_{\text{ccw}}(p,q) \leq d_{\text{ccw}}(r,s)$ and $d_{\text{ccw}}(q,r) \leq d_{\text{ccw}}(s,p)$ imply $|C_{\text{ccw}}(p,q)| \leq |C_{\text{cw}}(p,q)|$, since $d_{\text{ccw}}(p,q) \leq d_{\text{ccw}}(r,s) \leq d_{\text{ccw}}(q,p)$, and $|C_{\text{ccw}}(p,r)| \leq |C_{\text{cw}}(p,r)|$, since $d_{\text{ccw}}(p,r) = d_{\text{ccw}}(p,q) + d_{\text{ccw}}(q,r) \leq d_{\text{ccw}}(r,s) + d_{\text{ccw}}(s,p) = d_{\text{ccw}}(r,p)$. Second, the red split cycle shrinks when moving $q$ to $r$, i.e., $|C_{\text{cw}}(p,r)| \leq |C_{\text{cw}}(p,q)|$, due to the triangle inequality and the fact that it remains on the same side of the red shortcut.

The above implies that $\text{diam}(C + pr + rs) \leq \text{diam}(C + pq + rs)$, because of the following. The value of $\text{diam}(C + pq + rs)$ is the maximum of four values, viz., the largest distance between any two points on each of the four candidate diametral cycles (excluding $C$, since $pq$ and $rs$ are useful). When replacing $pq$ with $pr$ we decrease or maintain each of these four values whose maximum becomes $\text{diam}(C + pr + rs)$, since $pr$ and $rs$ are useful shortcuts.    ◂

## 3.2   Balancing Diametral Cycles

We show that every cycle has an optimal pair of alternating shortcuts where the bowtie and the hourglass are both diametral and we show that every convex cycle has an optimal pair of shortcuts where both split cycles are diametral, as well. We obtain these results by applying a sequence of operations that each slide the shortcuts along the cycle in a way that reduces or maintains the continuous diameter and brings the candidate diametral cycles closer to the desired balance. The last two operations only reduce the diameter for convex cycles as the shortcuts might get stuck at reflex vertices, which leads to our characterization of optimal shortcuts for convex and non-convex cycles.

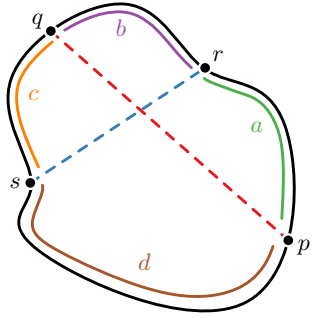

**Figure 19** The sections of a cycle with a pair of alternating shortcuts.

Let $pq$ and $rs$ be two alternating shortcuts and let $a = d_{\text{ccw}}(p, r)$, $b = d_{\text{ccw}}(r, q)$, $c = d_{\text{ccw}}(q, s)$, and $d = d_{\text{ccw}}(s, p)$. As depicted in Figure 19, we assume that the red split cycle contains $s$ and the blue split cycle contains $p$, i.e., $a+b \leq c+d$ and $b+c \leq a+d$. Furthermore, we abbreviate the lengths of the bowtie ($\bowtie$), the hourglass ($\mathbb{X}$), the red split ($\oslash$), and the blue split ($\oslash$) as follows.

$$\bowtie := a + c + |pq| + |rs| \quad \oslash := c + d + |pq| \quad \mathbb{X} := b + d + |pq| + |rs| \quad \oslash := a + d + |rs|$$

▶ **Lemma 3.6.** *For each relation $\sim \in \{<, =, >\}$ we have*

$$\bowtie \sim \mathbb{X} \iff a + c \sim b + d \qquad \oslash \sim \oslash \iff c + |pq| \sim a + |rs|$$
$$\bowtie \sim \oslash \iff a + |rs| \sim d \qquad \mathbb{X} \sim \oslash \iff b + |rs| \sim c$$
$$\bowtie \sim \oslash \iff c + |pq| \sim d \qquad \mathbb{X} \sim \oslash \iff b + |pq| \sim a$$

*and $pq$ and $rs$ are useful if and only if $|pq| + |rs| < a + c$ and $|pq| + |rs| < b + d$.*

**Proof.** The claims follow from the definitions of $\bowtie$, $\mathbb{X}$, $\oslash$, and $\oslash$. ◀

▶ **Lemma 3.7.** *Consider a pair of useful alternating shortcuts where one of the split cycles evenly divides the cycle. Then this split cycle must have length at most $\bowtie$ or at most $\mathbb{X}$.*

**Proof.** Assume, for the sake of a contradiction, that we have a pair of useful shortcuts where $\oslash$ divides the cycle evenly, i.e., $a + b = c + d$, and where $\bowtie < \oslash$ and $\mathbb{X} < \oslash$.

Then we have $a + |rs| < d$ and $b + |rs| < c$, by Lemma 3.6, which yields

$$a + |rs| < d = a + b - c < a - |rs| \ ,$$

leading to the contradiction $|rs| < 0$. ◀

▶ **Lemma 3.8.** *There exists an optimal pair of shortcuts in alternating configuration such that none of the split cycles is the only diametral cycle.*

**Proof.** Suppose $pq$ and $rs$ are useful and $\oslash$ is sole diametral, i.e., $\bowtie < \oslash$, $\mathbb{X} < \oslash$, and $\oslash < \oslash$.

Then we have $b + |pq| < a$ and $c + |pq| - |rs| < a$. We move $r$ clockwise along the cycle until we arrive at some $r'$ where $b' + |pq| = a'$ or $c + |pq| - |r's| = a'$, as in Figure 20.

Changing $r$ to $r'$ leads to the following changes in the candidate diametral cycles.

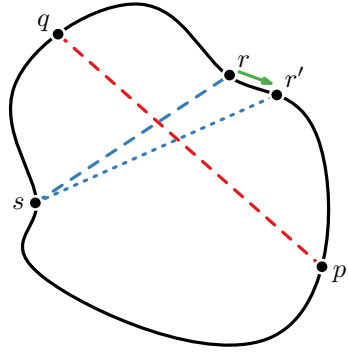

**Figure 20** Shrinking the blue split by moving $r$ clockwise to $r'$.

- The blue split shrinks or remains the same, i.e., $\oslash \geq \oslash'$.
- The red split remains the same, i.e., $\obslash = \obslash'$.
- The bowtie changes as the blue split does, i.e., $\bowtie \geq \bowtie'$ and $\bowtie' < \oslash'$.
- The hourglass remains the same or increases, i.e., $\mathbb{X} \leq \mathbb{X}'$.
- The hourglass increases when the blue split remains the same, i.e, $\oslash - \mathbb{X} > \oslash' - \mathbb{X}'$.

Consequently, $\mathrm{diam}(C+pq+r's) \leq \mathrm{diam}(C+pq+rs)$ and $\oslash' = \mathbb{X}'$ or $\oslash' = \obslash'$, which implies our claim, provided that $pq$ and $r's$ are useful shortcuts.

We argue that $r's$ is a shortcut and that $pq$ and $r's$ are useful. Assume, for the sake of a contradiction, that $r's$ is not a shortcut, i.e., $|r's| = d(r',s)$. Suppose the shortest path from $r'$ to $s$ in $C$ travels counter-clockwise around the cycle, i.e., $d(r',s) = d_{\mathrm{ccw}}(r',s)$. Then $s$, $r$, and $r'$ are colinear, since $r's$ contains $rs$. This contradicts $rs$ being a shortcut. Suppose, on the other hand, the shortest path from $r'$ to $s$ in $C$ travels clockwise around the cycle, i.e., $d(r',s) = d_{\mathrm{cw}}(r',s)$. Since $b+c \leq a+d$, the shortest path from $r$ to $s$ travels counter-clockwise along the cycle. This means that when moving $r$ to $r'$ we passed through a point $r''$ where the blue split $\oslash''$ evenly divides the cycle and is at least as large as the hourglass $\mathbb{X}''$ and strictly larger than the bowtie $\bowtie''$, contradicting Lemma 3.7. Since the assumption $|r's| = d(r',s)$ leads to a contradiction in both cases, $r's$ is a shortcut.

The shortcuts $pq$ and $r's$ remain useful, as $|pq| + |r's| < b' + d$ holds, since

$$|pq| + |r's| \leq |pq| + |rs| + d_{\mathrm{ccw}}(r',r) < b + d + d_{\mathrm{ccw}}(r',r) = b' + d \ ,$$

and, as $|pq| + |r's| < a' + c$ holds, since $|r's| < d(r',s) \leq b' + c$ and $b' + |pq| \leq a'$ imply

$$|pq| + |r's| < |pq| + b' + c \leq a' + c \ .$$

Therefore, there is a pair of optimal shortcuts where $\obslash$ is not the only diametral cycle. ◄

▶ **Lemma 3.9.** *There exists a pair of optimal shortcuts with $\bowtie = \mathbb{X}$.*

**Proof.** Suppose $pq$ and $rs$ are useful shortcuts with $\bowtie \neq \mathbb{X}$.

We balance $\bowtie$ and $\mathbb{X}$ using the following operations that maintain or decrease the continuous diameter while decreasing the difference between bowtie and hourglass. Two of these operations are illustrated in Figure 21.

1. As long as neither split cycle divides the cycle $C$ evenly, we shrink the larger split cycle in a way that decreases the difference of bowtie and hourglass:

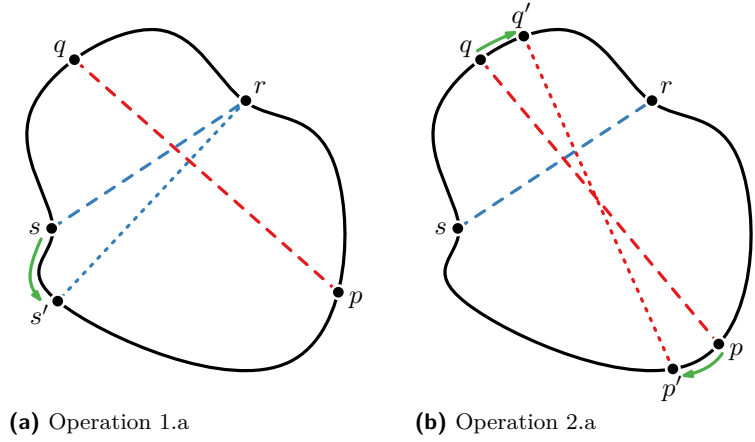

**(a)** Operation 1.a    **(b)** Operation 2.a

**Figure 21** Two of the operations used to balance bowtie and hourglass.

   **a.** When $\bowtie < \hourglass$ and $\oslash \leq \obslash$, we move $s$ counter-clockwise.
   **b.** When $\bowtie < \hourglass$ and $\oslash > \obslash$, we move $p$ clockwise.
   **c.** When $\bowtie > \hourglass$ and $\oslash \leq \obslash$, we move $r$ clockwise.
   **d.** When $\bowtie > \hourglass$ and $\oslash > \obslash$, we move $q$ counter-clockwise.
2. Once a split cycle evenly divides the cycle, we move the endpoints of the corresponding shortcut in the direction that decreases the difference between bowtie and hourglass:
   **a.** When $\bowtie < \hourglass$ and $\oslash$ evenly divides the cycle, we move $p$ and $q$ clockwise.
   **b.** When $\bowtie > \hourglass$ and $\oslash$ evenly divides the cycle, we move $p$ and $q$ counter-clockwise.
   **c.** When $\bowtie < \hourglass$ and $\obslash$ evenly divides the cycle, we move $s$ and $r$ counter-clockwise.
   **d.** When $\bowtie > \hourglass$ and $\obslash$ evenly divides the cycle, we move $s$ and $r$ clockwise.

For each operation, we argue that $pq$ and $rs$ remain useful shortcuts and that the diameter never increases while the difference between hourglass and bowtie always decreases.

Suppose $\bowtie < \hourglass$ and neither split is even. By repeatedly applying Operations 1.a and 1.b, we move $s$ counter-clockwise to $s'$ and we move $p$ counter-clockwise to $p'$ until $\bowtie' = \hourglass'$ or until one split is even. This causes the following changes in the candidate diametral cycles.

- Both splits shrink or remain the same, i.e., $\oslash \geq \oslash'$ and $\obslash \geq \obslash'$.
- The bowtie grows or remains the same, i.e., $\bowtie \leq \bowtie'$.
- The hourglass remains the same or shrinks, i.e., $\hourglass \geq \hourglass'$.

Altogether, this means $\mathrm{diam}(C + pq + rs) \geq \mathrm{diam}(C + p'q + rs')$, provided that $p'q$ and $rs'$ are useful shortcuts. Moreover, the bowtie grows when the hourglass remains the same and the hourglass shrinks when the bowtie remains the same, i.e., $\hourglass - \bowtie > \hourglass' - \bowtie'$.

Assume, for the sake of a contradiction, that $p'q$ is not a shortcut, i.e., $|p'q| = d(p', q)$. As the red split never switched sides during the operation, we have $d(p', q) = d_{\mathrm{ccw}}(p', q)$. This means that the line segment $p'q$ contains $pq$ contradicting our choice of $pq$ as a shortcut. Therefore, $p'q$ must be a shortcut. Symmetrically, we can argue that $rs'$ is a shortcut.

By the triangle inequality, we have $|p'q| \leq |pq| + d_{\mathrm{ccw}}(p', p)$ and $|rs'| \leq |rs| + d_{\mathrm{ccw}}(s, s')$. With $a' + c' \leq b' + d'$ from $\bowtie' \leq \hourglass'$, we obtain that $p'q$ and $rs'$ are useful, since

$$|p'q| + |rs'| \leq |pq| + d_{\mathrm{ccw}}(p', p) + |rs| + d_{\mathrm{ccw}}(s, s')$$
$$< a + d_{\mathrm{ccw}}(p', p) + c + d_{\mathrm{ccw}}(s, s') = a' + c' \leq b' + d' \ .$$

Applying Operations 1.a and 1.b, never increases the diameter while reducing the difference between bowtie and hourglass. Similarly, we can argue that the same holds for 1.c and 1.d.

Suppose $\bowtie < \Xi$ and $\oslash$ is an even split, i.e., $a + b = c + d$. Consider Operation 2.a. We simultaneously move $p$ clockwise to $p'$ and $q$ clockwise to $q'$ until $\bowtie' = \Xi'$. This causes the following changes in the candidate diametral cycles.

- The red split cycle remains an even split, i.e., $a' + b' = c' + d'$.
- The blue split cycle remains unchanged, i.e., $\oslash = \oslash'$.
- The bowtie grows or remains the same, i.e., $\bowtie \leq \bowtie'$.
- The hourglass remains the same or shrinks, i.e., $\Xi \geq \Xi'$.

Even though the red split might grow in length, it cannot determine the diameter, since $\oslash' \leq \bowtie'$ or $\oslash' \leq \Xi'$, due to Lemma 3.7. Altogether, this means $\text{diam}(C + pq + rs) \geq \text{diam}(C + p'q + rs')$, provided that $p'q'$ and $rs$ are useful shortcuts. Moreover, the bowtie grows when the hourglass remains the same and the hourglass shrinks when the bowtie remains the same, i.e., $\Xi - \bowtie > \Xi' - \bowtie'$, and, consequently, $\bowtie' = \Xi'$.

Assume, for the sake of a contradiction, that $p'q'$ is not a shortcut, i.e., $|p'q'| = d(p', q')$. Since $\oslash'$ is an even split, this would mean $|p'q'| = a' + b' = c' + d' = |C|/2$, i.e., the cycle $C$ is degenerate. Since we exclude degenerate cycles, $p'q'$ must be a shortcut.

We moved both $p$ and $q$ by the same distance $\Delta$ along the network. By the triangle inequality, we have $|p'q'| \leq |pq| + 2\Delta$. With $a' + c' = b' + d'$ from $\bowtie' = \Xi'$, this yields

$$|p'q'| + |rs| \leq |pq| + 2\Delta + |rs| < a + \Delta + c + \Delta = a' + c' = b' + d' \ ,$$

which implies that $p'q'$ and $rs$ are indeed useful. Therefore, Operation 2.a, never increases the diameter while reducing the difference between bowtie and hourglass until they are equal. Similarly, we can argue that the same holds for Operations 2.b, 2.c, and 2.e.

The above implies the claim, since, by using Operation 1 and Operation 2, we can transform every alternating configuration of useful shortcuts into another configuration where bowtie and hourglass have equal length without increasing the continuous diameter. ◀

▶ **Corollary 3.10.** *There exists a pair of optimal shortcuts that is in the alternating configuration such that none of the split cycles is the only diametral cycle and such that the bowtie and the hourglass have the same length.*

**Proof.** Let $pq$ and $rs$ be a pair of optimal shortcuts for a cycle $C$ where none of the splits is the only diametral cycle. Each of Operations 1.a, 1.b, 1.c, and 1.d from Lemma 3.9 shrinks the larger split cycle at the same rate as they shrink the larger of bowtie and hourglass. Thus, we do not create a sole diametral cycle by applying these operations. Furthermore, each of Operations 2.a, 2.b, 2.c, and 2.d rotates an even split that cannot become diametral by Lemma 3.7. By applying Lemma 3.9, we obtain a pair of optimal shortcuts $p'q'$ and $r's'$ with at least two diametral cycles and where bowtie and hourglass have the same length. ◀

▶ **Theorem 3.11.** *For every non-degenerate cycle, there exists an optimal pair of shortcuts such that the hourglass and the bowtie are both diametral.*

**Proof.** Let $C$ be a non-degenerate cycle. By Corollary 3.10, there is a pair of optimal shortcuts $pq$ and $rs$ where neither split cycle is the only diametral cycle and where $\bowtie = \Xi$.

Suppose that $\bowtie$ and $\Xi$ are not diametral. The cycle $C$ cannot be diametral, since $pq$ and $rs$ are useful. This means a split is diametral, i.e., $\oslash > \bowtie = \Xi$ or $\oslash > \bowtie = \Xi$. Since neither $\oslash$ nor $\oslash$ is the only diametral cycle, we have $\oslash = \oslash > \bowtie = \Xi$.

We shrink the splits by simultaneously moving $p$ and $r$ clockwise while moving $q$ and $s$ counter-clockwise, as illustrated in Figure 22. By moving $pq$ and $rs$ at appropriate speeds, we ensure that this operation maintains both the balance between the split cycles and the balance between bowtie and hourglass, i.e., $\oslash' = \varnothing'$ and $\bowtie' = \rtimes'$. This decreases the continuous diameter, provided that the line segments $p'q'$ and $r's'$ remain useful shortcuts.

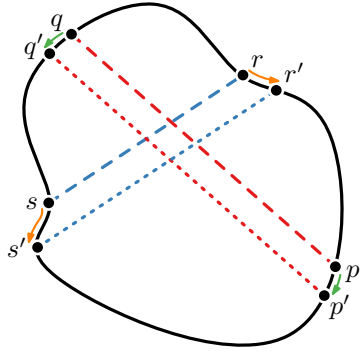

**Figure 22** Shifting the shortcuts to shrink the split cycles while maintaining the balance between both split cycles and the balance between bowtie and hourglass.

Assume, for the sake of a contradiction, that $p'q'$ is not a shortcut, i.e., $|p'q'| = d(p'q')$. By Lemma 3.7 we cannot pass through an even red split during our operation. Thus, we have $d(p', q') = d_{\mathrm{ccw}}(p', q')$, i.e., the line segment $p'q'$ contains $pq$ contradicting the choice of $pq$ as shortcut. Therefore, $p'q'$ is a shortcut. Symmetrically, we can argue that $r's'$ is a shortcut.

We argue that $p'q'$ and $r's'$ remain useful. From $\bowtie' = \rtimes' \leq \oslash' = \varnothing'$, we obtain $|p'q'| \leq d' - c'$, $|p'q'| \leq a' - b'$, and $|r's'| \leq c' - b'$, by Lemma 3.6. Together with $b' > 0$ and $a' + c' = b' + d'$, we derive that $p'q'$ and $r's'$ are useful, because

$$|p'q'| + |r's'| \leq d' - c' + c' - b' = d' - b' < d' + b' = a' + c' .$$

This means $p'q'$ and $r's'$ are useful shortcuts with $\bowtie' = \rtimes' = \oslash' = \varnothing'$ and $\mathrm{diam}(C+pq+rs) > \mathrm{diam}(C + p'q' + r's')$ contradicting the optimality of $pq$ and $rs$. Therefore, there exists a pair of optimal shortcuts where both the hourglass and the bowtie are diametral. ◀

▶ **Theorem 3.12.** *For every convex cycle, there exists an optimal pair of alternating shortcuts such that the hourglass, the bowtie, and the splits are diametral, i.e., $\bowtie = \rtimes = \oslash = \varnothing$.*

**Proof.** According to Theorem 3.11, there are optimal shortcuts $pq$ and $rs$ with $\bowtie = \rtimes \geq \oslash$ and $\bowtie = \rtimes \geq \varnothing$. Suppose we have $\bowtie = \rtimes > \oslash$ or $\bowtie = \rtimes > \varnothing$.

We establish the claim in three steps. First, we ensure that we can increase each split in a way that shrinks its shortcut. Second, we grow the smaller split until both splits are equal. Third, we grow both splits at the same rate until they are equal to bowtie and hourglass.

Moving $p$ counter-clockwise and $q$ clockwise *shifts* the red shortcut $pq$ in a way that increases the red split $\oslash$. As argued in Theorem 3.11, this shift maintains the balance between bowtie and hourglass and preserves usefulness.

- Suppose that shifting the red shortcut $pq$ to increase $\oslash$ shortens the red shortcut. Since the cycle is convex, the red shortcut shrinks as we continue to shift it.
- Suppose that shifting the red shortcut $pq$ to increase $\oslash$ increases the length of the red shortcut, as illustrated in Figure 23. Since the cycle is convex, the red shortcut shrinks as we shift the red split in the other direction, moving $p$ clockwise and $q$ counter-clockwise. We shift the red shortcut to shrink $\oslash$ until it becomes even and starts growing.

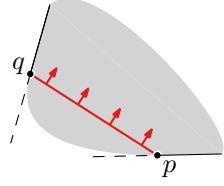

**Figure 23** The situation when shifting the red shortcut increases its length. Since the cycle (gray) is convex, shifting the red shortcut in the other direction must decrease its length.

Since each shift operation maintains $a + c$ and $b + d$, the bowtie and the hourglass change as $|pq| + |rs|$ changes. As argued above, we shift $pq$ or $rs$ in a way that decreases $|pq| + |rs|$ and thereby ⋈ and ⨉ while increasing the smaller split until both splits are equal. The same applies when we continue to shift $pq$ and $rs$ at the same time adjusting the speed of the shifts to maintain ◌ = ⊘. Eventually, we arrive at shortcuts $p'q'$ and $r's'$ with ⋈$'$ = ⨉$'$ = ◌$'$ = ⊘$'$ and $\mathrm{diam}(C + pq + rs) \geq \mathrm{diam}(C + p'q' + r's')$, since we only decreased the diametral cycles throughout the shift operations and maintain usefulness as argued in Theorem 3.11. ◀

▶ **Corollary 3.13.** *For every non-degenerate cycle, there exists an optimal pair of shortcuts such that the hourglass and the bowtie are diametral and such that each split cycle is diametral or the shortcut of the split has at least one endpoint at a reflex vertex.*

▶ **Corollary 3.14.** *For every convex cycle, there exists an optimal pair of shortcuts with $a + b \leq c + d$ and $b + c \leq a + d$ such that the following holds.*

$$a + c = b + d = \frac{|C|}{2} \tag{1}$$

$$|rs| = d - a = c - b \tag{2}$$

$$|pq| = d - c = a - b \tag{3}$$

$$d = \frac{|C|}{4} + \frac{|pq| + |rs|}{2} = \mathrm{diam}(C + pq + rs) \tag{4}$$

$$b = \frac{|C|}{4} - \frac{|pq| + |rs|}{2} = \mathrm{diam}(C) - \mathrm{diam}(C + pq + rs) \tag{5}$$

$$a = \frac{|C|}{4} + \frac{|pq| - |rs|}{2} \tag{6}$$

$$c = \frac{|C|}{4} + \frac{|rs| - |pq|}{2} \tag{7}$$

**Proof.** Equations (1), (2), and (3) follow from ⋈ = ⨉ = ◌ = ⊘ and Lemma 3.6. Equations (4)–(7) follow from ⨉ begin diametral, i.e., $\mathrm{diam}(C + pq + rs) = $ ⨉$/2$. ◀

## 4  A Linear-Time Algorithm for Convex Cycles

For convex cycles, we restrict our search to the pairs of shortcuts satisfying ⋈ = ⨉ = ◌ = ⊘, due to Theorem 3.12. We proceed as follows. First, we pick some point $p$ on the cycle $C$ and compute three points $q$, $r$, and $s$ such that ⋈ = ⨉ = ◌ = ⊘—regardless of whether $pq$ and $rs$ are shortcuts. We show that the points $q$, $r$, and $s$ exist and are unique for every point $p$ along $C$. Once we have balanced $p$, $q$, $r$, and $s$, we slide $p$ along $C$ maintaining the balance by moving $q$, $r$, and $s$ appropriately. We show that $q$, $r$, and $s$ move in the same direction as $p$ while preserving their order along $C$. Thus, each endpoint traverses each edge of the $n$ edges of $C$ at most once throughout this process, which therefore takes $O(n)$ time.

For the remainder of this section, we only focus on convex cycles with non-empty interior. Consider a cycle $C$ and a fixed point $p$ on $C$. We say a triple of points $q$, $r$, and $s$ is *in balanced configuration with $p$* when the points $p$, $r$, $q$, and $s$ appear counter-clockwise in this order along $C$, $d_{\text{ccw}}(p,q) \leq |C|/2$, $d_{\text{ccw}}(r,s) \leq |C|/2$, and $\bowtie = \mathbb{X} = \oslash = \obslash$.

▶ **Theorem 4.1.** *Consider a convex cycle $C$ and a point $p$ on $C$. There exists a triple $q$, $r$, $s$ of points on $C$ that are in balanced configuration with $p$.*

**Proof.** Suppose we place $s$ at some arbitrary position on $C$ with $|C|/4 \leq d_{\text{ccw}}(s,p) \leq |C|/2$. We have three objectives when placing $r$ and $q$. First, ensure $\bowtie = \mathbb{X}$ by enforcing a distance of $|C|/2 - d_{\text{ccw}}(s,p)$ between $r$ and $q$ along $C$. Second, place $r$ and $q$ such that $d_{\text{ccw}}(p,q) \leq |C|/2$ and $d_{\text{ccw}}(r,s) \leq |C|/2$. Third, ensure $\oslash = \obslash$ as follows:

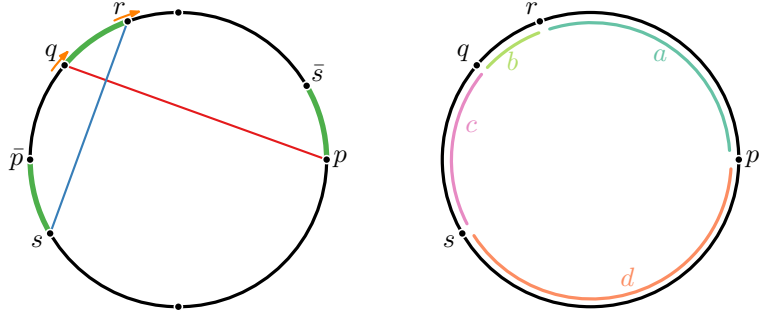

**Figure 24** Locating $r$ and $s$ to balanced the splits for fixed $p$ and $s$.

Let $\bar{p}$ and $\bar{s}$ be such that $d(p,\bar{p}) = d(s,\bar{s}) = |C|/2$. Suppose we slide $q$ and $r$ along the clockwise path from $\bar{p}$ to $\bar{s}$ while maintaining $d(q,r) = d(\bar{p},s)$, as illustrated in Figure 24. As $q$ moves clockwise $\oslash$ increases and $\obslash$ decreases. For convex cycles, at least one of $\oslash$ and $\obslash$ changes at any time during this motion. When $s$ is close enough to $\bar{p}$, we have $\oslash < \obslash$ when $q = \bar{p}$ and $\oslash > \obslash$ when $r = \bar{s}$. By the intermediate value theorem, there exist positions for $q$ and $r$ such that $\bowtie = \mathbb{X}$ and $\oslash = \obslash$ and these positions are unique, since $C$ is convex.

As $s$ moves closer to $p$, we reach a position $s^*$ where $\oslash = \obslash$ when $q = \bar{p}$ or where $\oslash = \obslash$ when $r = \bar{s}$. Suppose $s = s^*$ with $\oslash = \obslash$ for $q = \bar{p}$. Then $d = a = d(s,p)$ and $b = d(q,r) = |C|/2 - d = d_{\text{ccw}}(q,p) - d$. By Lemma 3.6, we have $c + |pq| = a + |rs|$, i.e.,

$$\bowtie = \mathbb{X} = b + d + |pq| + |rs| = d(q,p) - d + d + |pq| + |rs| > d(q,p) + |pq| = \oslash = \obslash \ .$$

Analogously, we obtain $\bowtie = \mathbb{X} > \oslash = \obslash$ when $s = s^*$ with $\oslash = \obslash$ for $r = \bar{s}$.

For all $s$ from $\bar{p}$ to $s^*$, the difference $\bowtie - \oslash$ is a continuous function of $d(s,p)$. We have $\bowtie = \mathbb{X} < \oslash = \obslash$ when $s = \bar{p}$, and we have $\bowtie = \mathbb{X} > \oslash = \obslash$, when $s = s^*$. By the intermediate value theorem there exists some position for $s$ such that $\bowtie = \mathbb{X} = \oslash = \obslash$. ◀

▶ **Corollary 4.2.** *Consider a convex cycle $C$ with $n$ vertices and a point $p$ on $C$. Suppose we know the counter-clockwise distance from $p$ to any other vertex. Then we can determine the triple of points on $C$ that are in balanced configuration with $p$ in $O(\log^2 n)$ time.*

**Proof.** Tracing the argumentation from the existence proof for Theorem 4.1, we perform a binary search for $s$ with a binary search for $q$ and $r$ in each step. The relations between the cycles $\bowtie$, $\mathbb{X}$, $\oslash$, and $\obslash$ will guide these binary searches, e.g., $q$ and $r$ need to move clockwise when $\oslash < \obslash$, and $s$ needs to move counter-clockwise when $\bowtie = \mathbb{X} > \oslash = \obslash$. Once we have identified the edges containing $q$, $r$, and $s$, we can express their exact positions as system of degree two polynomials and linear inequalities based on the conditions in Corollary 3.14. ◀

▶ **Lemma 4.3.** *Consider a convex cycle $C$. Suppose $p$ moves counter-clockwise along $C$. Then any three points in balanced configuration with $p$ are moving counter-clockwise, as well.*

**Proof.** Consider $p$ and $p'$ with $d_{\text{ccw}}(p, p') < |C|/4$. Let $q, r, s$ and $q', r', s'$ be triples of points on $C$ that are in balance with $p$ and $p'$, respectively. Note that $d_{\text{ccw}}(p, p') < |C|/4$ ensures that $p, s', q'$, and $r'$ appear clockwise in this order along $C$. Furthermore, let $\delta_x := d_{\text{ccw}}(p, x') - d_{\text{ccw}}(p, x)$ and $\Delta(xy) := |x'y'| - |xy|$ for any $x, y \in \{p, q, r, s\}$. By the triangle inequality, we have $\Delta(xy) \leq |\delta_x| + |\delta_y|$ where $\Delta(xy) = |\delta_x| + |\delta_y|$ occurs when the line segment $x'y'$ contains $xy$ and $\Delta(xy) = -|\delta_x| - |\delta_y|$ occurs when $xy$ contains $x'y'$.

We first argue that $q', r', s'$ lie counter-clockwise of $q, r, s$, respectively, i.e., $\delta_q, \delta_r, \delta_s \geq 0$, as illustrated in Figure 25. Then, we show $\delta_q, \delta_r, \delta_s > 0$, which means $q$, $r$, and $s$ move counter-clockwise as $p$ moves counter-clockwise and none of them remain at their position.

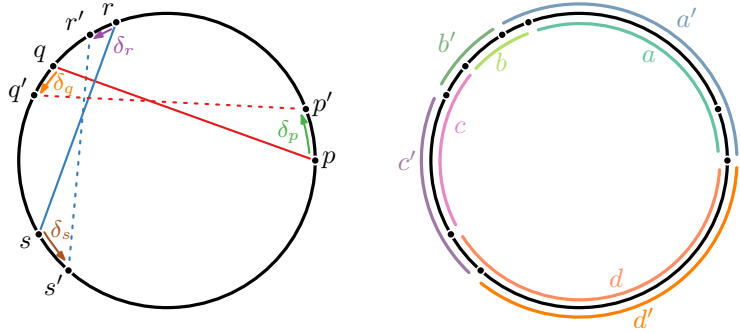

**Figure 25** Two balanced configurations $p, q, r, s$ and $p', q', r', s'$ with $\delta_p, \delta_q, \delta_r, \delta_s > 0$.

Regardless of the signs of $\delta_q$, $\delta_r$, and $\delta_s$, the following holds.

$$a' = a + \delta_r - \delta_p \qquad b' = b + \delta_q - \delta_r \qquad c' = c + \delta_s - \delta_q \qquad d' = d + \delta_p - \delta_s$$

Lemma 3.6 yields the following equations, since $\bowtie' = \rotatebox{90}{$\bowtie$}' = \oslash' = \oslash'$ and $\bowtie = \rotatebox{90}{$\bowtie$} = \oslash = \oslash$.

$$\bowtie' = \rotatebox{90}{$\bowtie$}' \wedge \bowtie = \rotatebox{90}{$\bowtie$} \iff \delta_p + \delta_q = \delta_r + \delta_s \tag{8}$$

$$\bowtie' = \oslash' \wedge \bowtie = \oslash \iff 2\delta_p = \delta_r + \delta_s + \Delta(rs) \tag{9}$$

$$\rotatebox{90}{$\bowtie$}' = \oslash' \wedge \rotatebox{90}{$\bowtie$} = \oslash \iff 2\delta_r = \delta_p + \delta_q + \Delta(pq) \tag{10}$$

Together with $\Delta(rs) \leq |\delta_r| + |\delta_s|$ we obtain

$$0 < 2\delta_p = \delta_r + \delta_s + \Delta(rs) \leq \delta_r + \delta_s + |\delta_r| + |\delta_s| \ . \tag{11}$$

This means that at least one of $\delta_r$ and $\delta_s$ must be positive. Assume, for the sake of a contradiction, that $\delta_s < 0$. Then $\delta_r > 0$ and Equation (11) implies $0 < \delta_p \leq \delta_r$. We distinguish the two cases $\delta_p < \delta_r$ and $\delta_p = \delta_r$. Suppose the former holds. Then Equation (10) and $\Delta(pq) = |p'q'| - |pq| \leq |\delta_p| + |\delta_q|$, yield $0 < \delta_r \leq \delta_p + \delta_q$, as

$$2\delta_p < 2\delta_r = \delta_p + \delta_q + \Delta(pq) \leq 2\delta_p + \delta_q + |\delta_q| \Rightarrow 0 < \delta_q + |\delta_q| \Rightarrow 0 < \delta_q \ ,$$

which contradicts $\delta_s < 0$, since Equation (8) implies $\delta_s = \delta_p + \delta_q - \delta_r \geq 0$. Now, consider the case when $\delta_p = \delta_r$. Note that $\delta_p > 0$ and $\delta_s < 0$ be assumption. Equation (8) implies $\delta_q = \delta_s < 0$. Revisiting Equations (9) and (10) with $\delta_p = \delta_r > 0$ and $\delta_q = \delta_s < 0$ yields

$$2\delta_p = \delta_r + \delta_s + \Delta(rs) \iff \Delta(rs) = \delta_r - \delta_s = |\delta_r| + |\delta_s| \tag{12}$$

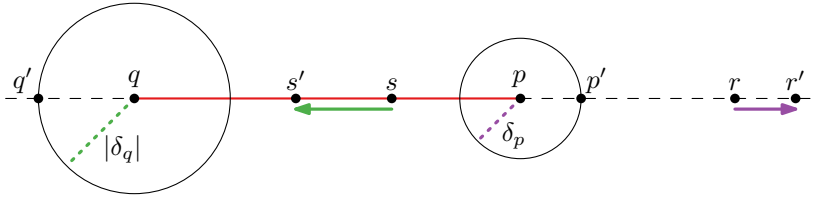

**Figure 26** The impossible constellation when $\Delta(rs) = |\delta_r| + |\delta_s|$ and $\Delta(pq) = |\delta_p| + |\delta_q|$ with $\delta_p = \delta_r > 0$ and $\delta_q = \delta_s < 0$ for a convex cycle. The cycle $C$ would be confined to a line.

$$2\delta_r = \delta_p + \delta_q + \Delta(pq) \iff \Delta(pq) = \delta_p - \delta_q = |\delta_p| + |\delta_q| \ . \tag{13}$$

Equation (12) means that $p'$ and $q'$ lie on the line $\ell(p,q)$ through $p$ and $q$, as illustrated in Figure 26. Since $C$ is assumed to be convex and, therefore, without self-intersections, the counter-clockwise path from $p'$ to $q'$ along $C$ coincides with the line segment $p'q'$. Both $s$ and $s'$ lie on $\ell(p,q)$, since $s$ lies on $pq$, since $q$ and $s$ travel clockwise for the same distance ($\delta_q = \delta_s < 0$), and since $p$ travels in the other direction ($\delta_p > 0$). Equation (13) means that $r'$ and $s'$ lie on the line $\ell(r,s)$ through $r$ and $s$. Since $ss'$ is contained in $p'q'$, the points $p$, $q$, $r$, $s$, $p'$, $q'$, $r'$, and $s'$ are colinear. Therefore, $C$ is confined to a line and cannot be convex.

Similarly, we can derive a contradiction from the assumption $\delta_r < 0$. Therefore, neither $\delta_r$ nor $\delta_s$ are negative and one of them is positive. In this case Equation (11) implies $\delta_p \leq \delta_r + \delta_s$ and we obtain $\delta_q = \delta_r + \delta_s - \delta_p \geq 0$ with Equation (8), i.e., $\delta_q, \delta_r, \delta_s \geq 0$ when $\delta_p > 0$.

Next, we argue that $\delta_q, \delta_r, \delta_s > 0$ when $\delta_p > 0$. Assume, for the sake of a contradiction, that $\delta_q = 0$. Equations (8) and (9) imply $\Delta(rs) = \delta_r + \delta_s$. This means that the line segment $r's'$ contains $rs$, as illustrated in Figure 27. Since both $r$ and $s$ are moving counter-clockwise along $C$, this implies that $C$ is confined to the line through $r$ and $s$, unless $\delta_r = 0$ or $\delta_s = 0$.

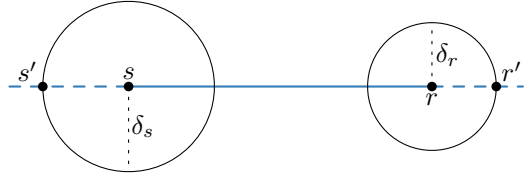

**Figure 27** The positions of $s'$ and $r'$ when $\delta_s > 0$, $\delta_r > 0$, and $|r's'| = |rs| + \delta_s + \delta_r$.

Suppose $\delta_r = 0$ and $\delta_s > 0$, i.e., $s$, $s'$, and $r = r'$ are colinear. Furthermore, we have $\delta_p = \delta_s > 0$, by Equation (8), and we have $\Delta(pq) = -\delta_p$ by Equation (10). Consequently, $p$, $p'$, and $q = q'$ are colinear. However, the cycle is confined to a line again, since $p$ and $p'$ lie on the line through $r$ and $s$. Similarly, we arrive at a contradiction when $\delta_r > 0$ and $\delta_s = 0$. Hence, we are only left with the option $\delta_r = \delta_s = 0$ when $\delta_p > 0$ and $\delta_q = 0$. Since this yields the contradiction $0 < \delta_p = \delta_r + \delta_s = 0$, we assert that $\delta_q > 0$ when $\delta_p > 0$.

Assume we have $\delta_p, \delta_q > 0$ and $\delta_r = 0$. By Equation (10), we have $\Delta(pq) = -\delta_p - \delta_q$ which means that $p'q'$ is strictly contained in the line segment $pq$ and, thus, $C$ is confined to the line through $p$ and $q$. Similarly, $C$ is confined to a line when $\delta_p, \delta_q, \delta_r > 0$ and $\delta_s = 0$.

As all other cases are exhausted, we have $\delta_r, \delta_q, \delta_s > 0$ when $\delta_p > 0$, i.e., when $p$ moves counter-clockwise along $C$ then so do any three points in balanced configuration with $p$.    ◀

▶ **Corollary 4.4.** *Consider a convex cycle $C$ and a point $p$ on $C$. There exists a unique triple $q$, $r$, $s$ of points on $C$ that are in balanced configuration with $p$.*

**Proof.** Assume, for the sake of a contradiction, that there are two distinct triples $q, r, s$ and $q', r', s'$ in balanced configuration with $p$. From the proof of Theorem 4.1, we already know that $q$ and $r$ are uniquely determined by $s$. Therefore, $s$ and $s'$ must be distinct.

Suppose $s$ lies closer to $p$ than $s'$, i.e., $d(s, p) < d(s', p)$. Invoking Lemma 4.3 with $s$ in the role of $p$ yields that $q$, $r$, and $p$ must move clockwise as $s$ moves clockwise to $s'$. Hence, $p, q', r'$ are not in balanced configuration for $s'$. This contradicts $q', r', s'$ being in balance with $p$. Therefore, $p$ has three unique points that are in balanced configuration with $p$. ◀

▶ **Theorem 4.5.** *Consider a convex cycle $C$ with $n$ vertices. We can compute an optimal pair of shortcuts for $C$ in $O(n)$ time.*

**Proof.** We pick an arbitrary point $p$ along some edge $e_p$ of $C$ and identify the edges $e_q$, $e_r$, and $e_s$ containing the points $q$, $r$, and $s$ that form a balanced configuration with $p$, as described in Corollary 4.2. We find a (locally) optimal pair of shortcuts $p^*q^*$ and $r^*s^*$ with whose endpoints lie on the edges $e_p, e_q, e_r, e_s$ by minimizing $d = \text{diam}(C + pq + rs)$ subject to $a + b \leq |C|/2$ and $b + c \leq |C|/2$, and the constraints stated in Corollary 3.14 that ensure $\bowtie\, =\, \bcancel{\bowtie}\, =\, \varotimes\, =\, \oslash$. Then, we identify the four edges that would host $p$, $q$, $r$, and $s$ next, if $p$ were to move counter-clockwise: for each endpoint $x \in \{p, q, r, s\}$, we calculate how far the other endpoints would move under the assumption that $x$ is the first point to hit a vertex. Corollary 4.4 guarantees that this calculation has a unique solution. Since all points move in the same direction as $p$, an edge $e$ will never host an endpoint $x$ in any subsequent step, once $x$ has left $e$. Therefore, the entire process takes $O(n)$ time. Since we encounter every four points in balanced configuration, we also encounter an optimal pair of shortcuts. ◀

## References

**1**  Victor Chepoi and Yann Vaxès. Augmenting trees to meet biconnectivity and diameter constraints. *Algorithmica*, 33(2):243–262, 2002.

**2**  Mohammad Farshi, Panos Giannopoulos, and Joachim Gudmundsson. Improving the stretch factor of a geometric network by edge augmentation. *SIAM Journal on Computing*, 38(1):226–240, 2008.

**3**  Fabrizio Frati, Serge Gaspers, Joachim Gudmundsson, and Luke Mathieson. Augmenting graphs to minimize the diameter. *Algorithmica*, 72(4):995–1010, 2015.

**4**  Yong Gao, Donovan R. Hare, and James Nastos. The parametric complexity of graph diameter augmentation. *Discrete Applied Mathematics*, 161(10-11):1626–1631, 2013.

**5**  Ulrike Große, Joachim Gudmundsson, Christian Knauer, Michiel Smid, and Fabian Stehn. Fast algorithms for diameter-optimally augmenting paths. In *42nd International Colloquium on Automata, Languages, and Programming (ICALP 2015)*, pages 678–688, 2015.

**6**  Chung-Lun Li, S. Thomas McCormick, and David Simchi-Levi. On the minimum-cardinality-bounded-diameter and the bounded-cardinality-minimum-diameter edge addition problems. *Operations Research Letters*, 11(5):303–308, 1992.

**7**  Jun Luo and Christian Wulff-Nilsen. Computing best and worst shortcuts of graphs embedded in metric spaces. In *19th International Symposium on Algorithms and Computation (ISAAC 2008)*, pages 764–775, 2008.

**8**  Anneke A. Schoone, Hans L. Bodlaender, and Jan van Leeuwen. Diameter increase caused by edge deletion. *Journal of Graph Theory*, 11(3):409–427, 1987.